\documentclass[12pt]{article}
\input epsf.sty

\begin{document}        

\pagestyle{empty}
\renewcommand{\thefootnote}{\fnsymbol{footnote}}

\begin{flushright}
{\small
OREXP 99-04\\
October 1999\\}
\end{flushright}
 
\vspace{.8cm}

\begin{center}
{\bf\large   
Study of top-quark reconstruction with LCD Fast Simulation
\footnote{Work supported by
Department of Energy grant DE-FG03-96ER40969 (Oregon).}}

\vspace{1cm}
{\bf
Masako Iwasaki
} \\
\vskip 0.2cm
{\it
Department of Physics, University of Oregon, Eugene,
OR 97403
} \\
\vskip 0.6cm

\medskip
\end{center}
 
\vfill

\begin{center}
{\bf\large   Abstract }
\end{center}

We report 
the study of top-quark reconstruction in $e^+e^- \rightarrow t\bar{t}$
events at a 500 GeV linear collider using the LCD Fast Simulator.
The final states of 6 jets as well as 4 jets and lepton are used.
In order to reconstruct the jets, 
the performance of charged and neutral cluster separation 
are studied.
We compare top-quark reconstruction for the LCD Small and
LCD Large detectors, including the effect of varying the
calorimeter granularity.

\vfill

\begin{center} 
{\it Presented at the world-wide study of physics and detectors for
future linear colliders (LCWS 99), 
8 Apr - 5 May 1999, Sitges, Barcelona, Spain}
\end{center}

\newpage
 
%
\pagestyle{plain}

\section{Introduction}               

Top-quark physics is an important topic for future $e^+e^-$ linear collider experiments.
Due to its large mass\cite{CDF D0},  
the top quark may play a special role in particle physics.
And since the final states are complicated, top-quark events provide
an important performance benchmark for detector and event reconstruction performance.
Therefore, realistic detector simulation studies are important,
though many generator level studies have already shown the potential of 
a future $e^+e^-$ linear collider for detailed studies of 
top-quark properties.

In this report, we present a study of top-quark reconstruction 
in $e^+e^- \rightarrow t\bar{t}$ using the
Linear Collider Detector (LCD) Fast Simulation.
The expected signature for $t\bar{t}$ production is two $b$ quarks 
and two $W$ bosons in the final states. 
The two $W$ bosons decay into either $q\bar{q\prime}$ or $l\bar{\nu}$,
giving final states configurations of (i)
two $b$ jets and four jets from W's (45\%), (ii)two $b$ jets, 
two jets and one charged lepton
(44\%), or (iii) two $b$ jets and two charged leptons(11\%). 
We study cases (i) and (ii),
where case (i) gives six jets and case (ii) has four jets 
and one charged lepton in the final states.

Jets were reconstructed by combining charged track and 
neutral cluster information. For this to work,
we have to identify whether each cluster is neutral or charged.
We will show neutral and charged cluster separation performance
for different detector parameters.
From these jets, we reconstruct top-quark candidates and
present the resulting reconstruction performance.

\section{Monte-Carlo Simulation}
Monte-Carlo simulation events are generated using the PANDORA-PYTHIA 
program\cite{pandora-pythia}. 
PANDORA generates parton-level 4-momentum vectors 
and PYTHIA 6.1\cite{pythia} hadronizes the partons.
For $\tau$ leptons, TAUOLA\cite{tauola} is used.
We generated 30,000 $e^+e^- \rightarrow t\bar{t}$ events with a 
top-quark mass of 175 GeV and a beam energy of 250 GeV.
Beamstrahlung and initial state radiation are included.
No other final states were generated.

To simulate the detector, we used the LCD Fast Simulation.
Charged particles within the magnet field follow helical trajectories, 
and their momenta and positions are smeared.
Electrons, photons, and hadrons produce clusters in the electromagnetic 
(EM) and hadronic (HAD) calorimeters.
Here, one cluster was made from one particle.
Energies and positions of clusters were smeared.

There are two detector models presently under study by the LCD group, a 
Small and a Large detector. In this study, we consider these two detector designs.
These models differ in geometry, materials, 
and resolutions. The detector parameters are summarized in Table.~\ref{table1}.
The smearing implemented in the LCD Fast Simulator is based on 
the resolutions.
For the calorimeters, we assume transverse position resolutions of 1cm/$\sqrt{E}$ 
(electrons and photons) or 5cm/$\sqrt{E}$ (hadrons). In order to 
consider the detector granularity, clusters within calorimeter cells
are merged, as described in Section 4.
Charged tracks with E$_{tracks} > 100$ MeV and 
$|\cos\theta|<0.90$ ($<$0.83 for Large), and clusters 
with E$_{cluster} > 100$ MeV and $|\cos\theta|<0.90$
were used.
\begin{table}[tbh]
\begin{center}
\begin{tabular}{l|c c}
\hline \hline
 & Small & Large \\
\hline
Vertex Detector & CCD & CCD \\
\,\,\,\,\,\,\,\, Impact parameter resolution & 
$4.5 \mu \bigoplus 5.5 \mu / p \sin^{2/3}$ &
$10.0 \mu \bigoplus 30.0 \mu / p \sin^{2/3}$ \\
\hline
Central Tracking & Si micro strips & TPC \\
\,\,\,\,\,\,\,\, Momentum resolution (High)& 
$\delta / P^2 \sim 6 \times 10^{-5}$ &
$\delta / P^2 \sim 5 \times 10^{-5}$ \\
\,\,\,\,\,\,\,\, \hspace{4cm} (Low)& 
$\delta P / P \sim 0.4\%$ &
$\delta P / P \sim 0.1\%$ \\
\hline
Electromagnetic Calorimeter & W/Si pads & Pb/scintillator\\
\,\,\,\,\,\,\,\, Barrel Inner Radius &  75 cm & 200 cm \\
\,\,\,\,\,\,\,\, Endcap Inner Z      & 150 cm & 300 cm \\
\,\,\,\,\,\,\,\, Energy resolution &
$\delta E / E \sim 12\%/ \sqrt{E} + 1 \%$ &
$\delta E / E \sim 15\%/ \sqrt{E} + 1 \%$ \\
\,\,\,\,\,\,\,\, Granularity & 20 mrad & 20 mrad \\ 
\hline
Hadron Calorimeter & Cu/scintillator & Pb/scintillator\\
\,\,\,\,\,\,\,\, Barrel Inner Radius & 140 cm & 250 cm \\
\,\,\,\,\,\,\,\, Endcap Inner Z      & 186 cm & 350 cm \\
\,\,\,\,\,\,\,\, Energy resolution &
$\delta E / E \sim 50\%/ \sqrt{E} + 2 \%$ &
$\delta E / E \sim 40\%/ \sqrt{E} + 2 \%$ \\
\,\,\,\,\,\,\,\, Granularity & 60 mrad & 60 mrad \\ 
\hline
Coil Magnet & & \\
\,\,\,\,\,\,\,\, Magnetic field & 6 Tesla & 3 Tesla \\
\,\,\,\,\,\,\,\, Inner Radius & 100 cm (outside EM Cal) & 
376 cm (outside HAD Cal) \\
\hline \hline
\end{tabular}
\end{center}
\caption{ Detector parameters for LCD Small and Large detectors.}
\label{table1}
\end{table}

\section{Charged and neutral cluster separation} 
Charged hadrons and electrons are detected
by both the tracker and the calorimeters.
At the center of mass energy of 500 GeV, $t\bar{t}$ events
have an average charged particle energy of about 2.4 GeV.
For these relatively low energies, the tracker resolution is better 
than the calorimeters. Therefore we use the tracker information 
for the charged particles, and remove the corresponding and the charged clusters.
The tracks are then combined with the neutral clusters to form jets. This
procedure is sometimes called the ``energy flow'' algorithm.

\begin{figure}
\epsfysize8cm
\hskip0.5in
\epsfbox{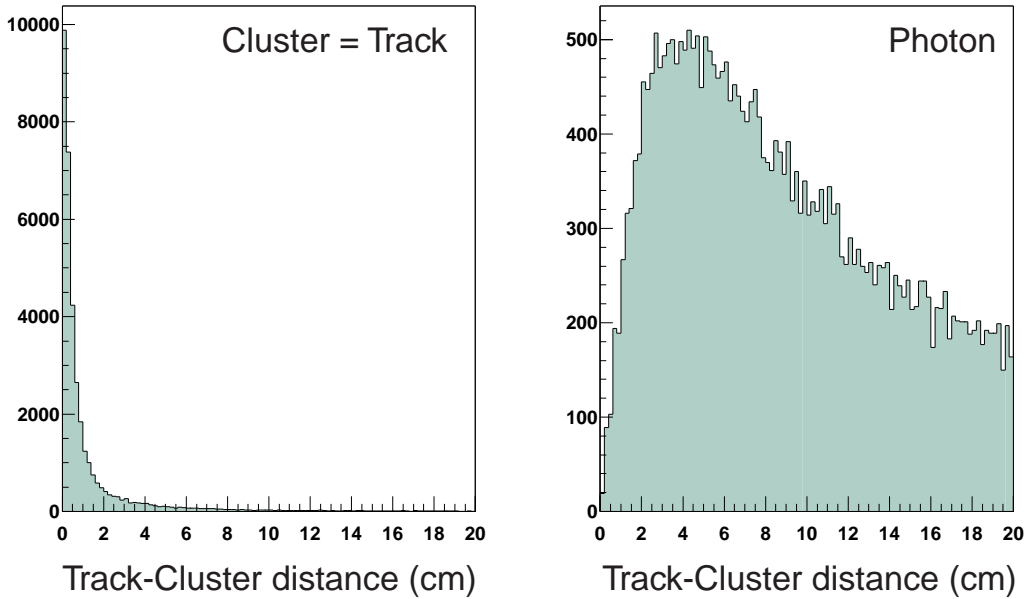}
\caption{ Distance between clusters and the associated tracks for
both clusters and tracks are from the same particle (left) 
and for photon clusters (right). 
}
\label{fig:matching}
\end{figure}

This technique requires a tight association between tracks and
the corresponding charged clusters. 
All charged tracks are extrapolated to the point where the cylindrical
radius is the same as the cluster radius.
Then the nearest track is associated with each cluster. 
Figure \ref{fig:matching} shows this distance when
both cluster and track result 
from the same particle (left), and for photon clusters (right). 
There is a peak around zero when the clusters and tracks are from the same
particle. On the other hand, there is no such a peak for the neutral
clusters. By applying a cut on this distance, charged
clusters were rejected.

\begin{figure}[tbh]
\begin{center}
\epsfysize8cm
\hskip0.5in
\epsfbox{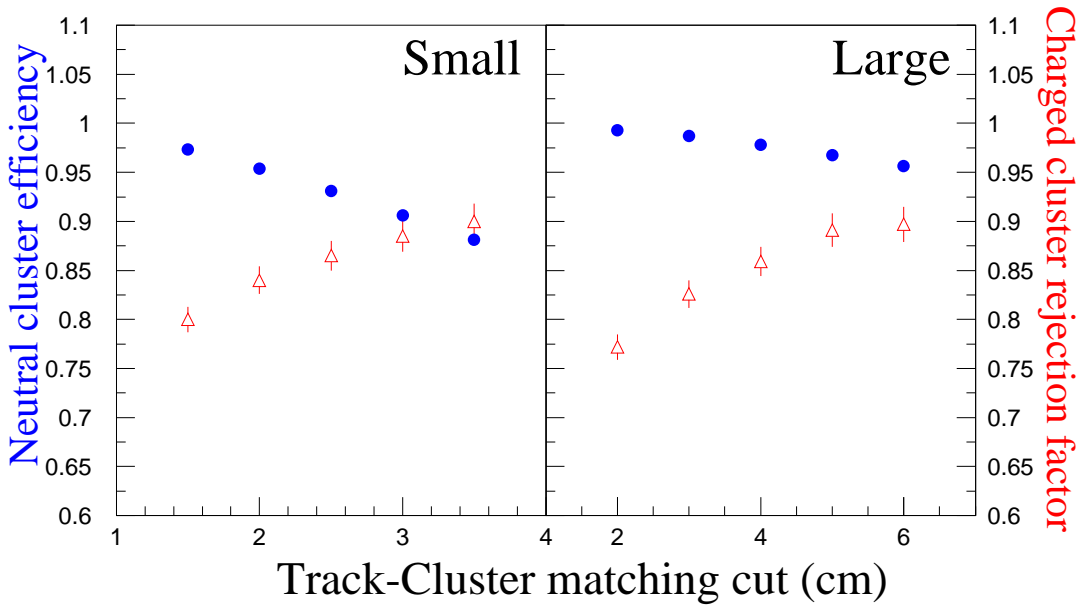}
\end{center}
\caption{Charged cluster rejection factors (solid circles) 
and neutral cluster efficiencies (open triangles) as 
functions of track-cluster matching cut values, for Small and
Large detectors.
}
\label{fig:efficiency}
\end{figure}

Figure \ref{fig:efficiency} shows charged cluster rejection factors (open
triangles) and 
neutral cluster efficiencies (solid circles) as a function of the value
of the track-cluster distance cut. 
The performance of charged and neutral cluster separation is proportional
to the magnetic field and the square of the calorimeter radius.
Therefore, the Large detector with a 3 Tesla magnetic field and a radius of
200 cm gives better separation than the Small detector with
6 Tesla field and 75 cm radius.
We used cut values of 2.5 cm and 4.0 cm for Small and Large detectors,
respectively. This gave a
87\% (86\%) charged cluster rejection factor
and a 93\%(98\%) neutral cluster efficiency for the Small
(Large) detector. 

\section{Cluster merging}
The LCD Fast Simulator makes one cluster from each particle
in the calorimeters.
To simulate the detector granularity and cluster width, which is
typically a few units of Moliere radius, 
we merge the clusters when the angular separation between
clusters is less than $\theta_{max}$, 
where $\theta_{max}$ is a measure of detector granularity.

Table~\ref{table2} shows the probability for a cluster to be merged 
with another cluster. The nominal granularity assigned to the
LCD Small and Large detector models is 20 mrad for each.
The 13 mrad case for the Small detector corresponds to a 1cm segmentation
at the EM Calorimeter surface. The 30 mrad granularity is the same 
as that of the  JLC calorimeter design.
For our analysis, we used the same granularity for EM and hadronic calorimeters,
and we did not use any longitudinal segmentation.
Therefore, clusters within $\theta_{max}$ are merged even though these clusters
might have different longitudinal positions.

\begin{table}[tbh]
\begin{center}
\caption{ Probability for a cluster to be merged.
The definition of $\theta_{max}$ is given in the text.}
\begin{tabular}{l|c c c}
\hline \hline
 & $\theta_{max}$ = 13 mrad & 20 mrad & 30 mrad \\
\hline
Small & 2.7 $\pm$ 0.4 \% & 5.4 $\pm$ 0.4 \% & 10.0 $\pm$ 0.5 \% \\   
Large &                  & 5.1 $\pm$ 0.4 \% &  9.4 $\pm$ 0.4 \% \\   
\hline \hline
\end{tabular}
\end{center}
\label{table2}
\end{table}

\section{Study of top-quark reconstruction}
\subsection{Reconstruction of $t\bar{t}\rightarrow$ 6 jets}
In this analysis, we used only $t\bar{t}\rightarrow$ 6 jet events 
(14000 events).
In order to reconstruct the $t\bar{t}\rightarrow$ 6 jets events, we select
events where the number of charged tracks is $\geq$ 30
and the visible energy exceeds 100 GeV. (Visible energy is calculated
using charged tracks and neutral clusters.)
To find jets, charged tracks and neutral
clusters are grouped into jets, using an invariant-mass
(JADE) algorithm.
First we apply $Y_{\rm cut} = 0.004$ and select events 
which have 6 or more jets. Then the $Y_{\rm cut}$ value is increased,
if necessary, until the event has exactly 6 jets.
The efficiency of this selection  is 71\% for $t\bar{t} \rightarrow$ 
6 jets events.

Since $t\rightarrow Wb$, $b$-quark tagging may be 
important for reducing background.
In order to tag the $b$-quark jet, we used the $N_{\rm sig}$ method, 
where $N_{\rm sig}$ indicates the number of tracks which have 3-D impact 
parameter with significance $> 3 \sigma$ (excluding $V^0$ decay tracks). 
A jet with $N_{sig}\geq 4$ was regarded as a $b$-jet candidate.
The method provides 87\% purity and 67\% efficiency
for $b$-quark jets.
After selecting $b$-quark jets, we form $W$ candidates by 
combining all remaining jet pairs.
Jet pairs with invariant mass within 12 GeV of 
the nominal W mass were kept. Top-quark candidates are then formed
from these $b$ and $W$ jets. To reduce random combinatoric background,
we require the quantity $x_E \equiv E_{3jets}/E_{beam}$ satisfy 
the condition $0.95 < x_E < 1.05$.
The remaining combinations with invariant mass in the range 
165 GeV to 185 GeV are regarded as top-quark candidates.
 
\begin{figure}[tbh]
\begin{center}
\epsfysize8cm
\hskip0.5in
\epsfbox{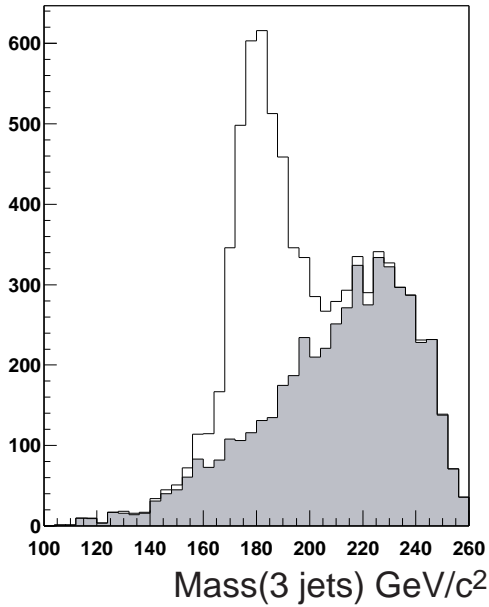}
\end{center}
\caption{Reconstructed mass distribution for top-quark signal
(open histogram) and random combinatoric background (shaded  histogram).
}
\label{fig:topmass}
\end{figure} 
Fig.~\ref{fig:topmass} shows the reconstructed top-quark mass distribution
for the Small detector.  
The reconstruction performance for Small and Large is 
summarized in Table~\ref{table3}, where the ``cluster merging
size'' is just $\theta_{max}$.  20 mrad corresponds to the granularity of the 
current specifications for both Small and Large detectors. A merging size of
13 mrad corresponds to a
segmentation of 1 cm at the inner radius of the Small EM calorimeter.
The difference in reconstruction performance between 13 mrad and 
20 mrad is very small (within 2\%). 
With 30 mrad, we obtain about 10\% smaller efficiency and 
a few percent worse angular and mass resolutions. 
This corresponds to the granularity of the JLC calorimeter.
The angular resolution
for reconstructed top is important for anomalous couplings analyses. Here,
it is determined from reconstructed top relative to MC truth.

Comparing the LCD Small and Large detectors, they have similar mass
resolution, but Large has about 10\% better angular resolution, and a few percent 
higher efficiency. This is apparently due to its larger $BR^2$, which gives
the Large detector an advantage for neutral and charged cluster separation,
as described in the previous section.
\begin{table}
\begin{center}
\caption{ Top-quark reconstruction performance for $t\bar{t}\rightarrow 6$
jets, as a function of
cluster merging size, for Small and Large detectors.}
\begin{tabular}{r|l|c c c}
\hline \hline
   & Cluster merging size & 13 mrad & 20 mrad & 30 mrad \\
\hline
Small & Top-quark candidates    & 
      2326 & 2303  & 2202 \\
      & Top-quark signal        & 
      1772 & 1742 & 1633 \\
      & Mass resolution (GeV)   & 
      9.59 $\pm$ 0.18 & 9.66 $\pm$ 0.21 & 9.86 $\pm$ 0.26 \\
      & Angular resolution (mrad)& 
      62.9 $\pm$ 1.2 & 63.8 $\pm$ 1.3 & 71.4 $\pm$ 1.5 \\
\hline
Large & Top-quark candidates    & 
      & 2469 & 2272 \\
      & Top-quark signal        &
      & 1899 & 1703 \\
      & Mass resolution (GeV)   &
      & 9.38 $\pm$ 0.17 & 9.96 $\pm$ 0.22 \\
      & Angular resolution (mrad)&
      & 56.7 $\pm$ 1.3 & 60.1 $\pm$ 1.0 \\
\hline \hline
\end{tabular}
\end{center}
\label{table3}
\end{table}

\section{Reconstruction of $t\bar{t}\rightarrow 4$ jets$ + $lepton}
In this study, we used 8600 $t\bar{t}\rightarrow 4$ jet$+ $lepton 
events in the generated $t\bar{t}$ Monte Carlo sample described above.
For event selection, we
require the number of charged tracks be $\geq 20$,
the visible energy exceed 100 GeV, and a lepton track (muon or electron)
with momentum $> 20$ GeV. No other criteria for lepton identification
are required.
All charged tracks, except for the charged lepton, 
are combined with the neutral clusters to form jets. As before, we 
first apply $Y_{cut} = 0.004$ to select events 
with 4 or more jets, then increase $Y_{cut}$ until the event has exactly 4 jets.

To tag the $b$-quark jets, we require $N_{sig} \geq 4$, resulting in
a 92\% purity and 64\% efficiency. 
We form the $W$ and top candidates exactly as was described for the
6-jet case. Of course, for this decay mode, we only have one $W$ and
one top which can be reconstructed in this way. 

The resulting reconstruction performance is summarized in
Table~\ref{table4}. As before, we find the
reconstruction performance for 13 mrad and 
20 mrad merging size to be within 2\%. 
For the 30 mrad case, we obtain 10\% (Small) and 5\% (Large) 
smaller efficiency and a few percent worse angular and mass resolution. 

Comparing the Small and Large detectors, the Large detector has similar mass
resolution and better angular resolution of 10\% (20 mrad) and 
2\% (30 mrad). On the other hand, the Small design has 10\% higher efficiency.
This is because for $t\bar{t} \rightarrow 4$ jet$ + $lepton reconstruction, 
both $b$-quark jet tagging and lepton identification are based on
tracker information, which has wider acceptance for the Small detector
($|\cos\theta|<0.90$), compared to the Large detector ( $<0.83$).

\begin{table}
\begin{center}
\caption{ Top-quark reconstruction performance in $t\bar{t}
      \rightarrow 4$ Jets$ + $lepton, as a function of the
cluster merging size, for Small and Large detectors.}
\begin{tabular}{r|l|c c c}
\hline \hline
   & Cluster merging size & 13 mrad & 20 mrad & 30 mrad \\
\hline
Small & Top-quark candidates    & 
      748  & 728 & 672 \\         
      & Top-quark signal        & 
      651  & 639 & 570 \\
      & Mass resolution (GeV)   & 
      9.61 $\pm$ 0.26 & 9.63 $\pm$ 0.23 & 9.64 $\pm$ 0.39 \\
      & Angular resolution (mrad)& 
      54.5 $\pm$ 1.6 & 53.0 $\pm$ 2.0 & 56.9 $\pm$ 2.1 \\ 
\hline
Large & Top-quark candidates    & 
      & 678 & 651 \\
      & Top-quark signal        &
      & 587 & 560 \\
      & Mass resolution (GeV)   &
      & 9.26 $\pm$ 0.36 & 9.18 $\pm$ 0.39 \\
      & Angular resolution (mrad)&
      & 47.5 $\pm$ 2.3 & 55.7 $\pm$ 2.3 \\
\hline \hline
\end{tabular}
\end{center}
\label{table4}
\end{table}

\section{Conclusion}
We have studied top-quark reconstruction in $e^+e^- \rightarrow t\bar{t}$
events at a 500 GeV linear collider using the LCD Fast Simulator. 
Two kinds of final states, six jets and 4 jets plus lepton, were used.
The performance of charged and neutral cluster separation and 
the effect of varying the calorimeter granularity were studied.

Comparing the Large and the Small detectors, 
Large has an advantage 
in the neutral and charged cluster separation, 
because of its larger calorimeter radius.
In the $t\bar{t}\rightarrow 6 jets$ reconstruction, 
the Large detector has 10\%
higher selection efficiency than Small due to this advantage.
On the other hand, the Small detector has 10\% higher efficiency in 
$t\bar{t} \rightarrow 4$ jets$ + $lepton due to its better
acceptance in $\cos\theta$ for charged tracks.

The top-quark reconstruction performance improves by only 2\% 
when the calorimeter granularity (simulated by merging clusters) is reduced
from 20 mrad to 13 mrad. When the granularity is increased to
30 mrad, we obtain about a 10\%  reduction in top reconstruction 
efficiency and a few percent worse angular and mass resolution. 

In these studies, we used the LCD Fast Simulator.
Since neutral and charged cluster separation is crucial
for top-quark reconstruction using this ``energy flow'' technique, 
more detailed and realistic studies, using full calorimeter shower simulations,
will be a necessary.



\end{document}